\documentclass[sigconf]{acmart}
\AtBeginDocument{%
  }

\copyrightyear{2025} 
\acmYear{2025} 
\setcopyright{cc}
\setcctype{by-nc}
\acmConference[WWW '25]{Proceedings of the ACM Web Conference 2025}{April 28-May 2, 2025}{Sydney, NSW, Australia}
\acmBooktitle{Proceedings of the ACM Web Conference 2025 (WWW '25), April 28-May 2, 2025, Sydney, NSW, Australia}
\acmDOI{10.1145/3696410.3714650}
\acmISBN{979-8-4007-1274-6/25/04}

\newtheorem{definition}{Definition}
\usepackage{multirow}
\usepackage{graphicx}
\usepackage{amsmath} 
\usepackage{algorithm}
\usepackage{algorithmic}
\usepackage{hyperref}

\begin{document}

\title[Domain-Informed Negative Sampling Strategies for Dynamic Graph Embedding]{Domain-Informed Negative Sampling Strategies for Dynamic Graph Embedding in Meme Stock-Related Social Networks}

\author{Yunming Hui}
\authornote{Corresponding author.}
\affiliation{%
  \institution{University of Amsterdam}
  \city{Amsterdam}
  \country{The Netherlands}}
\email{y.hui@uva.nl}

\author{Inez Maria Zwetsloot}
\affiliation{%
  \institution{University of Amsterdam}
  \city{Amsterdam}
  \country{The Netherlands}}
\email{i.m.zwetsloot@uva.nl}
  
\author{Simon Trimborn}
\affiliation{%
  \institution{University of Amsterdam}
  \department{Amsterdam School of Economics}
  \city{Amsterdam}
  \country{The Netherlands}}
\affiliation{%
  \institution{Tinbergen Institute}
  \city{Amsterdam}
  \country{The Netherlands}}
\email{s.trimborn@uva.nl}

\author{Stevan Rudinac}
\affiliation{%
  \institution{University of Amsterdam}
  \city{Amsterdam}
  \country{The Netherlands}}
\email{s.rudinac@uva.nl}



\begin{abstract}
Social network platforms like Reddit are increasingly impacting real-world economics. Meme stocks are a recent phenomena where price movements are driven by retail investors organizing themselves via social networks. To study the impact of social networks on meme stocks, the first step is to analyze these networks. Going forward, predicting meme stocks' returns would require to predict dynamic interactions first. This is different from conventional link prediction, frequently applied in e.g. recommendation systems. For this task, it is essential to predict more complex interaction dynamics, such as the exact timing. These are crucial for linking the network to meme stock price movements. Dynamic graph embedding (DGE) has recently emerged as a promising approach for modeling dynamic graph-structured data. However, current negative sampling strategies, an important component of DGE, are designed for conventional dynamic link prediction and do not capture the specific patterns present in meme stock-related social networks. This limits the training and evaluation of DGE models in such social networks. To overcome this drawback, we propose novel negative sampling strategies based on the analysis of real meme stock-related social networks and financial knowledge. Our experiments show that the proposed negative sampling strategies can better evaluate and train DGE models targeted at meme stock-related social networks compared to existing baselines.
\let\thefootnote\relax\footnote{The code is available at \href{https://github.com/YunmingHui/DINS}{https://github.com/YunmingHui/DINS}}
\end{abstract}




\maketitle

\section{Introduction} \label{sec_intro}

Social networks are playing an ever increasing role in the society~\cite{rao2022study,bakshy2012role}. Various studies show that social networks, such as Twitter~\cite{gu2020informational} and Reddit~\cite{umar2021tale}, also influence financial markets. The research presented in this paper is motivated by Reddit and the GameStop (GME) market frenzy occurring around January 2021 when users on the subreddit `Wallstreetbets' discussed GME and collectively caused a market frenzy~\cite{trimborn2024reddit}. It has become clear that internet users are a notable group influencing stock prices specifically for so called `meme stocks'~\cite{costola2021mementum}, stocks that receive significant attention on social media. To study the relationship between retail investors on social networks such as Reddit and the stock markets, a thorough understanding is needed of the network structure and behavior of the people posting on these networks. A first step to study the relationship is in understanding changes in posting behavior over time which may trigger stock market actions~\cite{trimborn2024reddit}. For this purpose, a dynamic network model which captures the dynamics of posting behavior at the individual node level is needed. 

Dynamic graph embedding (DGE) has emerged as an effective tool for tackling these challenges~\cite{xue2022dynamic,xie2020survey}. Graphs naturally describe social networks by representing individuals as nodes and their interactions as edges, providing a structured framework for analysis. DGE builds on this by transforming the nodes and edges into continuous vector representations (node embeddings), preserving both the network's structural and temporal properties. This approach allows DGE to capture the dynamic evolution of social networks over time, enabling insights into complex user interactions and facilitating predictions of network behavior.

For DGE, dynamic link prediction (DLP) is an important component which predicts if there is a link between two nodes based on their embeddings~\cite{barros2021survey}. From a technical perspective, DLP can evaluate the quality of generated embeddings and serve as the training objective. From an application perspective,  predicting when two users will interact in the future based on embeddings can help identify stock market trends~\cite{zhang2018exploiting,lei2022use,chen2013study}, such as renewed interest in a stock, which may manifest in a new stock market frenzy. 

DGE models need to accurately predict both existing interactions (positive samples) and nonexistent connections (negative samples)~\cite{leskovec2010predicting,daud2020applications}. If the model only predicts that all interactions exist, it may achieve good performance on positive samples but will incorrectly identify nonexistent connections, leading to unreliable and misleading results. However, in meme stock-related social networks, the number of negative samples far exceeds that of positive samples. Due to the huge size of these social networks, users typically only communicate with a fraction of other users~\cite{wang2019silent,gonzalez2010structure}. Therefore, the majority of these negative samples provide little valuable information, as many users may never interact. We conjecture that using such obvious non-connections for model training and evaluation will focus the models' prediction ability upon these obvious non-connections, whereas the real challenge lies in predicting negative samples which are difficult to predict in real social networks. This highlights the need to carefully select informative negative samples, a process known as negative sampling~\cite{yang2020understanding}.

Most existing negative sampling strategies (NSSs) for DLP are primarily based on random or heuristic approaches~\cite{chen2024curriculum,poursafaei2022towards,poursafaei2023exhaustive}. 
For instance, random negative sampling is one of the most widely used strategies~\cite{poursafaei2022towards}. It generates one negative sample for each positive sample $(u,v,t)$, where $u,v,t$ is the sender, receiver, and occurrence time of the interaction, by replacing $v$ with a random user. Such a NSS samples many obvious non-connections, resulting in deceivingly outstanding performance. 

State-of-the-art (SOTA) DGE models can achieve the AUC (Area Under the Receiver Operating Characteristic Curve) over 0.9 on certain datasets when trained and evaluated using the random sampling~\cite{rossi2020temporal,zhang2023tiger,gao2022novel}. However, the practical use for real applications is low. For illustration, consider the use case of predicting when the users who have already interacted (i.e., there are edges connecting two nodes) will interact again. This is important because repeated interactions often indicate renewed interest or users' joint and repeated interest in a stock, which can lead to price movements for meme stocks~\cite{trimborn2024reddit}. To evaluate the model's prediction ability in such a case, we generated three types of negative samples for each positive sample $(u, v, t)$: $(u, v, t + 6h)$, $(u, v, t + 12h)$, and $(u, v, t + 24h)$. These samples test whether the model can correctly predict if nodes that have interacted will interact again after 6, 12 and 24 hours. We use the dataset AMC (cf. Section~\ref{sec:dataset}) and the SOTA DGE model Temporal Graph Networks (TGNs)~\cite{rossi2020temporal} as an example. The results in Table~\ref{tab_instruction} show that the TGNs achieved an AUC of 0.9736 when trained and tested using random negative sampling, closely matching results reported in the original paper~\cite{rossi2020temporal}. However, the performance dropped strongly when tested with the other three types of negative samples. This indicates that random sampling limits the ability of TGNs to predict when previously interacting nodes will interact again.

\begin{table}[ht]
\caption{Test AUC of TGNs with various NSSs (AMC Dataset, January for training and February for validation and testing).}
\label{tab_instruction}
\begin{tabular}{ccccc}
\hline
NSS & Random Sampling & 6h     & 12h    & 24h    \\ \hline
Test AUC & 0.9736 & 0.6041 & 0.6982 & 0.7681 \\ \hline
\end{tabular}%
\end{table}

This example shows that the design of NSSs should be closely tied to domain knowledge which has also been established by \cite{ma2023exploring}. In the settings like meme stock-related social networks, interactions between users are not random or uniform. A generic NSS may miss important information, leading to suboptimal performance in predictive tasks. By incorporating domain-specific knowledge, such as understanding the significance of predicting the exact time of repeated interactions, a more effective NSS can be developed. 

In this paper, we analyzed three real-world meme stock-related social network datasets containing interactions on Reddit related to three companies, GameStop (GME), American Multi-Cinema (AMC), and BlackBerry (BB). We identified several key characteristics of meme stock-related social networks, such as the frequency of interactions between users, and the presence of unique interaction types such as loops. Based on these insights, we developed several individual NSSs specifically tailored to these network properties. Each strategy captures a distinct aspect of the network dynamics. We also developed a joint NSS, incorporating these individual NSSs. To overcome the imbalance between positive and negative samples in the training set, which may lead to a performance drop, we implemented positive enhancement that includes additional positive samples to balance positive and negative samples. 

In summary, we make the following contributions:

(1) We explored the application of DGE models to meme stock-related social networks and found that the current design of NSSs, an important component of DGE models, limits the performance of DGE models in this kind of social networks.

(2) We proposed several individual NSSs based on the analysis of three real-world meme stock-related social networks and corresponding financial domain knowledge. Each of them evaluates a certain part of DGE models' prediction ability in meme stock-related social networks. We also proposed a NSS named \textbf{D}(omain)\textbf{I}(nformed)\textbf{N}(egative)\textbf{S}(ampling) that combines these individual strategies and balances positive and negative samples.

(3) We conducted extensive experiments to show the effect of NSSs in the evaluation and training of DGE models. The experimental results show that proposed NSSs can improve DGE models' prediction ability in meme stock-related social networks.

\section{Preliminaries}

In this section, we define several important concepts and introduce the relationship between them.

\subsection{Dynamic Social Network Representation}
As we discussed in Introduction, dynamic social networks can be represented by the dynamic graph. In DGE, the dynamic graph can be either continuous or discrete. The continuous dynamic graph provides higher time resolution, allowing a more accurate representation of the network evolution~\cite{xue2022dynamic,lee2019temporal}. Hence, we represent dynamic social networks with the continuous dynamic graph.

\begin{definition}[Continuous dynamic graph]
A continuous dynamic graph is denoted by $\mathcal{G}=(\mathcal{V}, \mathcal{E})$, where $\mathcal{V}$ is the node set containing $n$ nodes and $\mathcal{E} = \{(u_i, v_i, t_i) \mid 1 \leq i \leq m \}$ ($0\leq t_1 \leq \cdots \leq t_m=T$) is the edge set containing $m$ directed edges. The source node, destination node, and the timestamp of an edge $e_i\in \mathcal{E}$ are denoted by $u_i\in \mathcal{V}$, $v_i\in \mathcal{V}$, and $t_i$ respectively. $T$ is the latest timestamp of the observed period.
\end{definition}

\subsection{DGE, DLP and NSS}
In this subsection, we define DGE, DLP and NSS formally. The relationship between them is also introduced. 

\begin{definition}[Dynamic Graph Embedding]
A dynamic graph embedding model is a function that maps a dynamic graph $\mathcal{G} = (\mathcal{V}, \mathcal{E})$ to a time-dependent continuous vector space. It assigns each node $u \in \mathcal{V}$ a time-specific embedding $\mathbf{z}_u(t) \in \mathbb{R}^d$, where $d$ is the dimension of the embedding. The node embeddings should preserve the evolving relationships and interactions between nodes over time.
\end{definition}

\begin{definition}[Dynamic Link Prediction]  
Given a dynamic graph $\mathcal{G} = (\mathcal{V}, \mathcal{E})$, dynamic link prediction aims to predict whether a future edge $(u, v, t)$ ($u\in \mathcal{V}$, $v\in \mathcal{V}$ and $t > T$) exits based on $\mathcal{G}$.
\end{definition}

\begin{definition}[Negative Sampling Strategy]
Given a continuous dynamic graph $\mathcal{G} = (\mathcal{V}, \mathcal{E})$, the set of all possible negative samples is defined as: $\mathcal{E}_{\text{neg}} = \{(u_i, v_i, t_i) \mid u_i, v_i \in \mathcal{V}, (u_i, v_i, t_i) \notin \mathcal{E}\}$. A negative sampling strategy is a method selecting a subset $\mathcal{E}_{\text{neg}}' \subseteq \mathcal{E}_{\text{neg}}$ for training or evaluation.
\end{definition}

DLP can be done based on embeddings as DGE models capture the evolving relationships between nodes in embeddings. NSS provides informative samples for DGE to learn better embeddings and for DLP to make more accurate predictions.

\section{Related Work} \label{sec_realted_work}

In this section, we review DGE methods and analyze existing NSSs.

\subsection{Dynamic Graph Embedding Models}
According to the survey of Barros et al.~\cite{barros2021survey}, learning-based DGE models have become the dominant approach in the field today. Thus, we focus on these learning-based models in this paper. 

Learning-based DGE models leverage deep learning techniques, such as recurrent neural networks (RNNs)~\cite{elman1990finding}, graph neural networks (GNNs)~\cite{scarselli2008graph}, and attention mechanisms~\cite{vaswani2017attention}, to capture the evolving relationships in dynamic networks. One prominent category of these models incorporates memory mechanisms which store and update node-specific information over time to better capture temporal dependencies~\cite{trivedi2019dyrep,rossi2020temporal,wang2021apan,ma2020streaming}. In addition to memory-based models, other approaches leverage advanced architectures like transformers, such as DyGFormer~\cite{yu2023towards} and GraphERT~\cite{beladev2023graphert}. Cong et al. claim that complex neural networks such as RNNs and attention mechanism are not always necessary, and proposed GraphMixer that relies on multi-layer perceptrons (MLPs)~\cite{cong2023we}. The majority of these DGE models use DLP as one of the tasks to evaluate the quality of generated embeddings \cite{trivedi2019dyrep,rossi2020temporal,wang2021apan,ma2020streaming,yu2023towards,beladev2023graphert}. In addition, many models use DLP as the learning objective too \cite{rossi2020temporal,cong2023we,yu2023towards,wang2021apan}.

\subsection{Negative Sampling Strategy in DLP}
The most common NSS used is Random Negative Sampling~\cite{poursafaei2022towards}. To generate negative samples, the destination node $v$ of each positive sample $(u,v,t)$ is replaced with a random node selected from all nodes. Random Negative Sampling is employed by the majority of studies developing DGE models. Though it is a straight-forward method to implement, the generated negative samples are mostly uninformative because the two nodes are likely to have never interacted before and therefore have completely different embeddings.

Recent works suggest that better NSSs are needed for DLP. Poursafaei et al.~\cite{poursafaei2022towards} argue that two nodes may connect multiple times. To address this, they proposed Historical Negative Sampling. They generate negative samples $(u',v',t')$ by requiring that node $u'$ and $v'$ have been connected at some
time before $t'$. These negative samples do provide more information compared to random sampling, but they still focus on one specific aspect of the network.

Some studies~\cite{chen2024curriculum,gao2024towards} use the idea of curriculum learning. They first generate all negative samples. Then, they select more difficult negative samples as model training progresses according to a criteria they defined for measuring the difficulty of negative samples. 
When generating negative samples, they replace $v$ of each positive sample $(u,v,t)$ with all nodes except nodes $u$ and $v$. If a node never or rarely becomes a source node, its relationship with other nodes is then not well captured by these negative samples. When designing the difficulty measurement criteria, they lack the consideration of domain knowledge.
The NSS proposed by Poursafaei et al.~\cite{poursafaei2023exhaustive} also considers all negative samples and reduces the number of negative samples by merging some samples occurring close together in time. However, they do not consider the situation that some nodes never or rarely act as a source node.

Overall, existing NSSs show the following deficiencies: 1) Sample many node pairs that never interacted, providing little useful information, or only focus on previously connected nodes, ignoring relationships between nodes that have never interacted; 2) Ignore the fact that nodes in social networks can act as both sender and receiver and that these roles often interchange; 3) Most approaches fail to incorporate domain knowledge, which could optimize the sampling process for specific network characteristics. Thus, these negative samples are not well suited for evaluating or training DGE models used for predicting meme stock-related social networks.

To address these deficiencies, we analyse three real-world meme stock-related social networks in combination with financial domain knowledge, to design NSSs that can better model these networks.

\section{Datasets} \label{sec:dataset}

We study three meme stock-related social network datasets~\cite{wang2024social} collected from WallStreetBets (WSB), shown as r/wallstreetbets on Reddit, which is a financial community where participants discuss investments. These three datasets include interactions regarding three companies: GameStop (GME), American Multi-Cinema (AMC), and BlackBerry (BB). The stock prices of all three companies were strongly influenced by these interactions during the time when these interactions happened~\cite{mancini2022self,trimborn2024reddit,chacon2023will}.

Reddit uses a post-comment structure, where user behavior falls into two categories. First, a user can create a post. Second, a user can comment on an existing post. Therefore, in the original datasets, users are treated as nodes ($\mathcal{V}$) and their interactions form the directed edges ($\mathcal{E}$). When a user creates a post, this is represented as a loop, forming an edge from the user to themselves $(u, u, t_i)$. When user $u$ comments on a post made by user $v$, this creates a directed edge from node $u$ to node $v$, denoted as $(u, v, t_i)$, where $t_i$ is the timestamp of the interaction in UTC format.  

We adjusted the original dataset as follows: 1) Removed unknown users and excluded data from months with excessively sparse interactions; 2) Reduced the time resolution of these datasets to 5 minutes. This means that all timestamps within each 5-minute interval were grouped and assigned the same timestamp. The descriptive statistics of the processed datasets are presented in Table~\ref{tab:statis-datasets}. 

\begin{table}[ht]
\caption{Descriptive statistics of three datasets. Unique node pairs refer to interactions between two distinct users, where the direction of the interaction matters. Loops represent edges whose source and destination node are the same.}
\label{tab:statis-datasets}
\resizebox{\columnwidth}{!}{%
\begin{tabular}{ccccccc}
\hline
Dataset & \begin{tabular}[c]{@{}c@{}}Nodes\\ $|\mathcal{V}|$\end{tabular} & \begin{tabular}[c]{@{}c@{}}Edges\\ $|\mathcal{E}|$\end{tabular} & \begin{tabular}[c]{@{}c@{}}Unique \\ Node Pairs\end{tabular}  & Loops                                                      & Start Date & End Date   \\ \hline
GME     & 517,975                                                       & 3,976,267                                                     & \begin{tabular}[c]{@{}c@{}}2,692,485\\ (67.71\%)\end{tabular} & \begin{tabular}[c]{@{}c@{}}134,010\\ (3.37\%)\end{tabular} & 2020-09-01 & 2021-08-31 \\
AMC     & 313,006                                                       & 2,207,981                                                     & \begin{tabular}[c]{@{}c@{}}1,544,006\\ (69.92\%)\end{tabular} & \begin{tabular}[c]{@{}c@{}}192,917\\ (8.73\%)\end{tabular} & 2021-01-01 & 2021-12-31 \\
BB      & 104,453                                                       & 406,916                                                       & \begin{tabular}[c]{@{}c@{}}305,349\\ (75.03\%)\end{tabular}   & \begin{tabular}[c]{@{}c@{}}30,434\\ (4.47\%)\end{tabular}  & 2021-01-01 & 2021-12-31 \\ \hline
\end{tabular}%
}
\end{table}

\section{Methodology} \label{section:method}
\begin{figure}[t]
    \centering
    \includegraphics[width=\columnwidth]{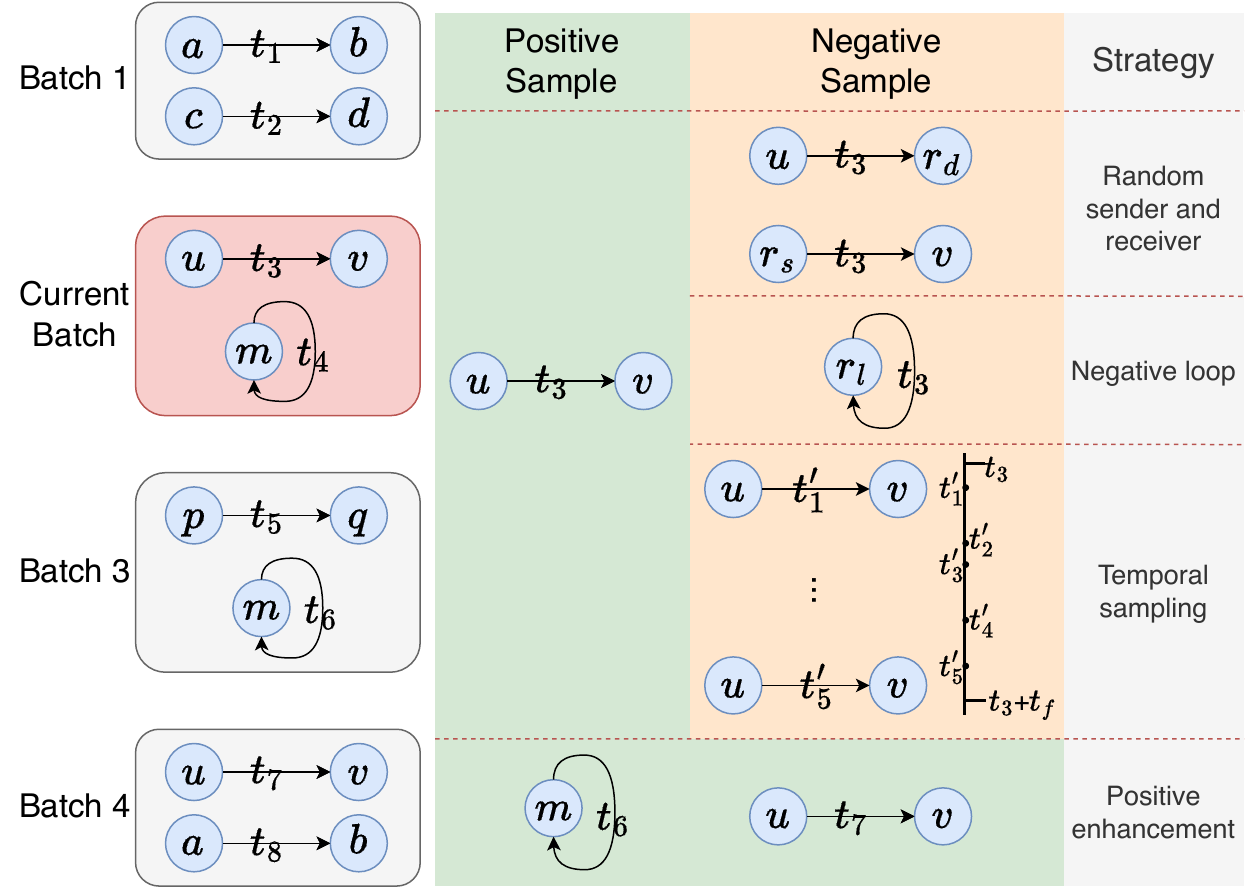}
    \caption{Visualization of proposed NSSs. Taking a dynamic network with 8 interactions as an example. The batch size is 2, meaning that the interactions are processed in groups of two. The showed negative samples are generated according to positive sample $(u,v,t_3)$.}
    \label{fig:example-negativ-samples}
\end{figure}

In this section, we first propose three individual NSSs. Each strategy captures a particular aspect of the network and is designed based on specific characteristics of meme stock-related social networks as well as financial domain knowledge. Then we propose a joint NSS that combines these individual strategies and balances the positive and negative samples. These strategies are visualized in Figure~\ref{fig:example-negativ-samples}.

In practical applications, learning-based DGE models typically divide the edge set $\mathcal{E}$ into multiple batches, containing a fixed number of edges. As shown in Figure~\ref{fig:example-negativ-samples}, all interactions within a batch are processed at once, with each batch being handled sequentially. This allows for more efficient computation and better memory management, especially when dealing with large datasets. Consequently, negative sampling is also performed on a per-batch basis. Thus, for illustration, we consider negative sampling on a batch $\mathcal{E}_{batch} = \{(u_i, v_i, t_i) \mid b \leq i < b+k \}$ containing $k$ interactions in the $b$-th batch. We additionally define $\mathcal{ET}_{batch} = \{t_i \mid b \leq i < b+k \}$ containing all timestamps in $\mathcal{E}_{batch}$.

\subsection{Individual Negative Sampling Strategies}

\subsubsection{Random sender and receiver}

We begin with the most fundamental evaluation, predicting the relationship between two nodes. In this NSS, we do not strictly test the model's ability to predict when two nodes will interact, but to evaluate whether the model can accurately predict if an interaction will happen.

The random NSS~\cite{poursafaei2022towards}, introduced in Section~\ref{sec_realted_work}, also serves this purpose and works in bipartite datasets, where source nodes represent users and destination nodes represent items, such as products or services. While effective for predicting the next item a user might be interested in, this method falls short in the context of meme stock-related social networks, where the relationships between any two nodes are of interest~\cite{haq2022short}. In such networks, a node can act as both a source node and a destination node, and these roles often interchange frequently. 

To solve this problem, we generate negative samples by randomly replacing both the source and destination nodes of each positive sample. Specifically, for each positive sample $(u,v,t) \in \mathcal{E}_{batch}$, two negative samples are generated: $(u,r_d,t)$ and $(r_s,v,t)$ where both $r_d$ and $r_s$ are randomly selected from $\mathcal{V} \setminus \{u,v\}$.

This strategy addresses the shortcomings of random negative sampling by enhancing the model's ability to evaluate the potential for interactions between any two nodes in a network, regardless of their typical roles as source or destination. By this the model predicts the next likely interaction based on meme stock-related social network dynamics. If these interactions occur in bulk, it may spill over into meme stock market activity~\cite{trimborn2024reddit}.

\subsubsection{Temporal sampling}
Temporal sampling additionally tests the model's ability to predict the exact time when two nodes will interact. As shown in Table~\ref{tab:statis-datasets}, the unique node pairs constitute only 67.71\%, 69.96\% and 75.03\% of the total number of edges in GME, AMC and BB, respectively. This shows that a large number of node pairs interacted more than once, while others interact only once. 
In meme stock-related social networks, accurately predicting when users who have already interacted will interact again is crucial. This is because repeated interactions often signal renewed interest or joint and sustained interest in a stock, which can influence price movements in meme stocks.

Temporal sampling works by generating negative samples for node pairs that have interacted in the past, but at future time points where no interaction has occurred. Specifically, for each positive sample $(u,v,t) \in \mathcal{E}_{batch}$, $q$ negative samples are generated: $\{(u, v, t_n) \mid 1 \leq n \leq q \}$ where $(u, v, t_n) \notin \mathcal{E}$ for any $n$, and the timestamps $t_n$ are uniformly and randomly distributed within the interval $[t, t+t_f]$. To adapt to different data sets, $q$ and $t_f$ are set as adjustable parameters. However, to avoid information leakage, $t+t_f$ should not be greater than the largest timestamp in $\mathcal{ET}_{batch}$ i.e., $t+t_f \leq \max(\mathcal{ET}_{batch})$.

Temporal sampling evaluates the model’s ability to predict interaction timing by introducing negative samples at future timestamps where no interaction has yet occurred. By focusing on previously interacting node pairs and generating future interactions that have not happened, the strategy tests the model's capability to identify true future interactions among potential ones.

\subsubsection{Negative loops}

As we discussed in Section~\ref{sec:dataset}, due to the post-comment structure used by Reddit, loops account for 3.37\%, 8.73\%, and 4.47\% of interactions in GME, AMC and BB dataset, respectively (cf. Table~\ref{tab:statis-datasets}). Predicting the existence of a loop is important because it may indicate that the user is re-engaging, potentially driven by new developments or shifts in stock performance.

This NSS is designed to evaluate if a DGE model can well predict the existence of loops. Specifically, for each timestamp $t \in \mathcal{ET}_{batch}$, one negative sample is generated: $(r_l,r_l,t) \notin \mathcal{E}$ where $r_l$ is a random node that has not formed a self-loop, i.e., $(r_l, r_l, t')$ does not exist for any $t' < t$. We require $r_l$ to be a node that has not formed a self-loop in order to specifically evaluate the model's ability to predict whether a node that is not expected to form a loop will indeed do so. 
The evaluation of nodes that are expected to form loops is already incorporated into the temporal sampling strategy.

\subsection{Combination}
The above individual NSSs each captures a specific aspect of the dynamic graph. In this subsection, we explain how they are combined to create a more comprehensive and effective NSS, named DINS, which is detailed in Algorithm~\ref{alg_negative_sampling}.

DLP is not only an evaluation tool but also serves as a crucial task during the training process for many DGE models. During the evaluation phase, all available negative samples can be used to fully assess the model's performance. However, in the training process, using too many negative samples can strongly skew the data distribution and create an imbalance~\cite{he2009learning,liu2021pick}. In particular, to properly evaluate whether the model can accurately predict the interaction time between two nodes, temporal sampling creates $q$ times more negative samples than positive samples. This could potentially cause the model to consistently predict no interaction between nodes to achieve lower loss values, thereby affecting the model's prediction accuracy.

To address this issue, DINS combines all these individual NSSs while maintaining a balance of positive and negative samples by implementing positive enhancement. Positive enhancement is conducted as follows (line 15-22 in Algorithm~\ref{alg_negative_sampling}): for each edge $(u,v,t)$ happening after the current batch, we check if nodes $u$ and $v$ interacted within the current batch. If so, the positive sample $(u,v,t)$ is added to the current sample set. However, the number of added positive samples does not exceed the size of the current batch to avoid increasing the overall training time or introducing unnecessary computational overhead. This ensures that the model learns to accurately distinguish between the presence and absence of edges, rather than defaulting to negative predictions. 

In Appendix~\ref{appendix:summary_NSSs}, we provide additional discussion on specific meme stock- related social network structures and how these related to the proposed DINS.

\begin{algorithm}[ht]
\caption{Negative Sampling Strategy for Training}
\label{alg_negative_sampling}
\begin{algorithmic}[1] 

\STATE \textbf{Input}: $\mathcal{E}_{batch}$, $\mathcal{V}$, $\mathcal{ET}_{batch}$, $t_f$, $q$, $k$

\STATE $\mathcal{S} \leftarrow \emptyset$  \COMMENT{Initialize a collection for all samples }

\FOR{$e=(u,v,t) \textbf{ in } \mathcal{E}_{batch}$}
    \STATE $r_s, r_d \leftarrow \textbf{random}(\mathcal{V} \backslash \{u,v\}),\textbf{random}(\mathcal{V} \backslash \{u,v\})$ 
    \STATE $\mathcal{S} \leftarrow \mathcal{S} \cup \{(r_s,v,t)\}$ \COMMENT{Random sender}
    \STATE $\mathcal{S} \leftarrow \mathcal{S} \cup \{(u,r_d,t)\}$ \COMMENT{Random receiver}

    \STATE $t_1, \cdots, t_q  \leftarrow \textbf{random}([t, t+t_f])$
    \STATE $\mathcal{S} \leftarrow \mathcal{S} \cup \{(u,v,t_1),\cdots,(u,v,t_q)\}$ \COMMENT{Temporal sampling}
    
\ENDFOR

\STATE $\mathcal{V}_l \leftarrow$ nodes have never formed a loop
\FOR{$t \textbf{ in } \mathcal{ET}_{batch}$}
    \STATE $r_l \leftarrow \textbf{random}(\mathcal{V}_l)$
    \STATE $\mathcal{S} \leftarrow \mathcal{S} \cup \{(r_l,r_l,t)\}$ \COMMENT{Negative loops}
\ENDFOR

\STATE $\mathcal{E}_{after} \leftarrow \{(u,v,t)|t> \max (\mathcal{ET}_{batch})\}$

\STATE positive\_count $\leftarrow$ 0
\FOR{$e=(u,v,t) \textbf{ in } \mathcal{E}_{after}$}
    \IF{$\exists t' s.t. (u,v,t')\in \mathcal{E}_{batch}$ \textbf{ and } positive\_count$< k$}
        \STATE $\mathcal{S} \leftarrow \mathcal{S} \cup \{(u,v,t)\}$ \COMMENT{Positive enhancement}
        \STATE positive\_count $\leftarrow$ positive\_count$+1$
    \ENDIF
\ENDFOR
\STATE \textbf{Output} $\mathcal{S}$
\end{algorithmic}
\end{algorithm}

\section{Experiments}
In this section, we validate our proposed individual NSSs and DINS with extensive experiments. All experiments are conducted on a machine with an Intel Xeon Platinum 8360Y (2.4 GHz, 18 cores), 128 GiB DDR4 RAM, and a NVIDIA A100 (40 GiB HBM2 memory), running Linux release 8.6.

\subsection{Dynamic Graph Embedding Models}
To demonstrate that our NSS can generalize and perform well regardless of the specific DGE model architecture, we selected three distinct DGE models with varying designs: 

1) \textbf{DyGFormer}~\cite{yu2023towards} employs advanced transformer architectures, which allows more sophisticated modeling of temporal dependencies in dynamic graphs. The authors of DyGFormer claim that it outperforms SOTA DGE models in various tasks, making it a strong candidate for evaluating the effectiveness of proposed NSSs. 

2) \textbf{GraphMixer}~\cite{cong2023we} achieves comparable or even superior performance using multi-layer perceptrons (MLP) instead of complex architectures used by other models, while also converging faster. This makes it an ideal choice for testing our NSS across different types of DGE models, allowing us to assess its effectiveness on simpler yet efficient models.

3) \textbf{TGNs}\cite{rossi2020temporal} is one of the most widely recognized DGE model that utilize memory mechanisms. TGNs has gained wide-spread attention due to its high performance on various datasets, making it a strong representative of memory-based DGE models. 

Dynamic graph embedding (DGE) can be either transductive or inductive~\cite{xu2020inductive}. Transductive learning limits predictions to nodes in the training set, while inductive learning allows the model to generalize and predict for unseen nodes. As we do not know beforehand whether a new user will be introduced to the network, we adopt the transductive approach in all experiments.

\subsection{Experimental Setting}

\subsubsection{Baselines}
We select two strategies as baselines.

\textbf{Random Negative Sampling (Random)} is widely used by most dynamic graph embedding studies~\cite{rossi2020temporal,cong2023we,yu2023towards,wang2021apan}. Choosing this as a baseline allows us to establish a standard for comparison, ensuring that our proposed methods are evaluated against a commonly accepted and effective approach.

\textbf{Historical Negative Sampling (Historical)}~\cite{poursafaei2022towards} is chosen because it has proven effective in capturing the temporal dynamics of node pairs that have interacted before. Its ability to leverage past interactions makes it a strong reference point for evaluating models, especially in dynamic environments where repeated interactions carry important signals.

We attempted to include curriculum learning based NSSs~\cite{chen2024curriculum,gao2024towards}, but their limited reproducibility based on the provided resources prevented incorporation in our experiments.

\subsubsection{Dataset split}
To avoid potential bias introduced by dataset splitting and to account for the varying interaction frequencies in meme stock-related social networks over different periods, we employed time series cross-validation. The implementation involved dividing the dataset by month, where each month's data was used for training, and the following month's data was used for validation and testing. For the GME dataset, due to the large volume of interactions in January, February, and March, we further subdivided these months. The specific method for dataset splitting and the data volume for each month after the split, can be found in Appendix~\ref{appendix:statistics_split_datasets}.

\subsubsection{Evaluation}
To evaluate the performance, we tested each model using seven different types of negative samples. The type Random Sender and Random Receiver are generated by NSS random sender and receiver. The type Loop are generated using NSS negative loop. Type 6h, 12h, and 24h are derived from temporal sampling, with $t_n$ fixed at 72, 144, and 288 timestamps, respectively, to assess the model's ability to predict relationships between previously interacting nodes over 6h, 12h, and 24h intervals. Type Overall includes all of the aforementioned negative samples. For each type, we used all positive samples in the test set along with the corresponding negative samples for the specific category.

The evaluation metric we selected is the AUC (Area Under the Receiver Operating Characteristic Curve)\cite{huang2005using}. We chose AUC because it provides a robust measure of a model’s ability to distinguish between positive and negative samples, regardless of class imbalance. AUC evaluates the trade-off between true positive and false positive rates, making it particularly suitable for our tasks where the ratio of positive to negative samples can vary significantly~\cite{ling2003auc}. AUC is also used by most DGE studies~\cite{xiong2019dyngraphgan, rossi2020temporal, cong2023we,yu2023towards}.

\subsubsection{Hyper-parameter Setting} The $q$ and $t_f$ for temporal sampling in our proposed NSS, DINS, is set to 5 and 288 timestamps (1 day) respectively for all experiments. The hyper-parameter settings of the three DGE models (see Appendix~\ref{appendix:hypersetting}) are based on the original paper and fine tuned on our datasets.

\subsection{Effect of Negative Sampling Strategy}

In this subsection, we analyse the effect of NSSs in both evaluating and training DGE models, respectively. We use three NSSs: random, historical, and DINS (proposed) to train three DGE models: TGNs, DyGFormer, and GraphMixer on three datasets, BB, AMC, and GME. This results in a total of 27 experiments.

\subsubsection{Evaluation}

\begin{figure*}[ht]
\centering
\includegraphics[width=\textwidth]{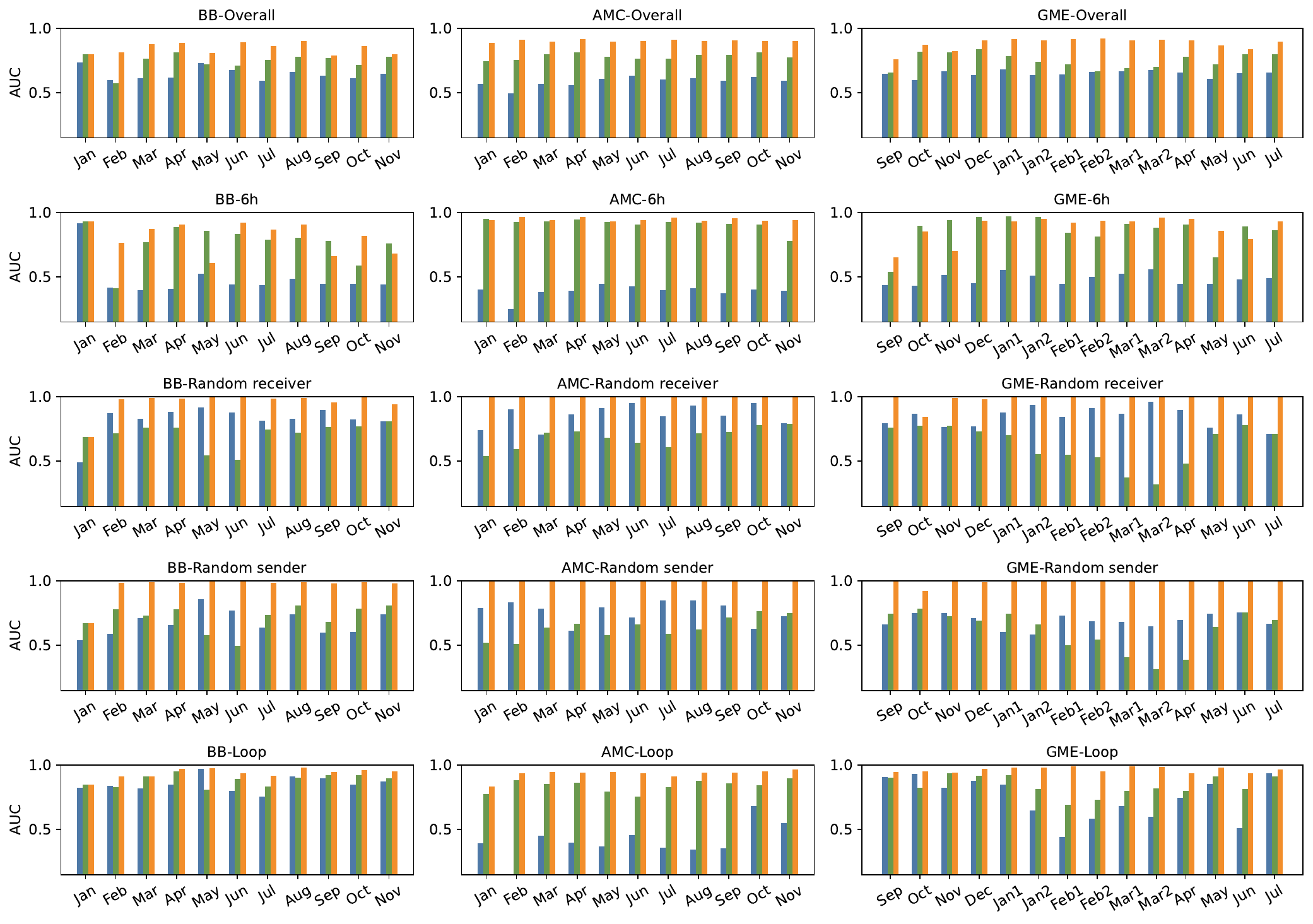}
\caption{Evaluation results of DyGFormer trained with three different NSSs. The \textcolor[HTML]{4E79A7}{blue bar}, \textcolor[HTML]{6A994E}{green bar} and \textcolor[HTML]{F28E2B}{orange bar} (from left to right) show the results of model trained with random NSS, historical NSS and DINS (proposed).}
\label{fig:time-series-DyGFormer}
\end{figure*}

We first show the effect of using different NSSs for evaluation on the same model (trained with the same NSS on the same dataset). In Figure~\ref{fig:time-series-DyGFormer}, we present the results of DyGFormer on the three datasets, showing how its performance varies with different NSSs used during evaluation. The results for the other two models can be found in the Appendix~\ref{appendix:results_tgn_graphmixer}. The experimental results are similar for all three DGE models. In Appendix~\ref{appendix:Robustness}, we provide a robustness analysis to better understand the effect of randomization in our experiments.

DyGFormer trained with random NSS (blue bars) achieves high AUC scores when evaluated with Random Receiver-type negative samples. However, when evaluated with 6h-type negative samples, the AUC falls below 0.5 in most cases, except for BB in January. Although AUC scores are a little bit higher when evaluated using other types of negative samples, the performance remains below that of Random receiver-type negative samples. For DyGFormer trained with historical NSS (green bars), the model shows higher AUC scores when evaluated with 6h-type negative samples, but similar to the previous case, AUC significantly drops when evaluated with other types of negative samples. DyGFormer trained with our proposed DINS (orange bars) performs well across most evaluation types except 6h-type on some months. This highlights the need for evaluation with different and diverse sampling strategies, anchored in domain knowledge.

Additionally, as we discussed in Section~\ref{section:method}, each of these proposed NSSs plays an important role in evaluating different aspects of meme stock-related social networks. Since the model's performance varies depending on the type of negative samples used for evaluation, our proposed NSSs provides a more comprehensive evaluation framework. This allows us to capture a wider range of interaction dynamics in meme stock-related social networks, ensuring that models are tested more thoroughly across multiple dimensions of the network's behavior.

\subsubsection{Training} 
Next, we show the effect of NSS used in training, starting again with the results on DyGformer. When evaluating with overall-type negative samples, the model trained using DINS consistently outperforms those trained with the two baseline strategies across all datasets. This suggests that incorporating our proposed strategy during training enhances the model's overall predictive ability for meme stock-related social networks. We then examine performance across various types of negative samples. When evaluated with 6h-type negative samples, the model trained using DINS shows slightly lower AUC than the model trained with historical NSS. This is expected, as the historical NSS focuses primarily on distinguishing this specific type of negative sample. However, for other types of negative samples, the model trained with DINS consistently demonstrates significantly better performance compared to those trained with historical negative sampling. When comparing to the model trained with random negative sampling, the model trained with DINS shows higher performance across all types of negative samples.

Following, we analyze the impact of NSSs on the training of all DGE models. The average ranking of each model and NSS combination across the monthly splits of each dataset is shown in~Table~\ref{tab:overall-result}. The results show that DINS consistently outperforms the other two baselines across all DGE models and datasets. Notably, the proposed strategy demonstrates a significant advantage with DyGFormer and TGNs, almost always achieving the top rank. Considering that these three models use different designs, we believe that proposed NSS can enhance the prediction ability of various DGE models on meme stock-related social networks.

\begin{table}[ht]
\caption{Overall rank of DGE models trained using different NSSs across three datasets. Bold indicates the best rank.}
\label{tab:overall-result}
\begin{tabular}{ccccc}
\hline
Strategy   & Dataset              & DyGFormer & GraphMixer & TGNs  \\ \hline
Random     & \multirow{3}{*}{BB}  & 2.81      & 2.63       & 3.00  \\
Historical &                      & 2.09      & 2.09       & 2.00  \\
DINS   &                      & \textbf{1.09}      & \textbf{1.27}       & \textbf{1.00}  \\ \hline
Random     & \multirow{3}{*}{AMC} & 3.00      & 2.54       & 1.90 \\
Historical &                      & 2.00      & 2.18       & 3.00  \\
DINS   &                      & \textbf{1.00}      & \textbf{1.27}       & \textbf{1.09}  \\ \hline
Random     & \multirow{3}{*}{GME} & 3.00      & 2.50       & 2.28  \\
Historical &                      & 2.00      & 2.07       & 2.57  \\
DINS   &                      & \textbf{1.00}      & \textbf{1.42}       & \textbf{1.14}  \\ \hline
\end{tabular}%
\end{table}

\subsection{Ablation Study}
In this subsection, we conducted ablation studies to validate the impact of each individual NSS on the performance of DGEs using the BB dataset. First, we remove each individual NSS from DINS and train the model. The performance is evaluated using overall-type negative samples. The results of DyGFormer are shown in Table~\ref{tab:ablation} and results of other two DGE models are shown in Appendix~\ref{appendix:ab}, showing similar results.

\begin{table*}[ht]
\caption{AUC of ablation studies for NSSs for DyGFormer on BB dataset.}
\label{tab:ablation}
\begin{tabular}{l|ccccccccccc}
\hline
Month  & Jan    & Feb    & Mar    & Apr    & May    & Jun    & July   & Aug    & Sep    & Oct    & Nov    \\ \hline
Proposed            & 0.7990 & 0.8135 & 0.8779 & 0.8872 & 0.8073 & 0.8917 & 0.8620 & 0.9016 & 0.7897 & 0.8644 & 0.7971 \\
- \ temporal   & 0.7721 & 0.8021 & 0.7952 & 0.7916 & 0.7921 & 0.7597 & 0.7813 & 0.7865 & 0.7646 & 0.7436 & 0.7617 \\
- \ loop     & 0.8620 & 0.8093 & 0.7971 & 0.8856 & 0.8063 & 0.8209 & 0.8371 & 0.8307 & 0.8520 & 0.8359 & 0.7846 \\
- \ sender   & 0.8968 & 0.9132 & 0.8276 & 0.8203 & 0.8552 & 0.7426 & 0.7262 & 0.7996 & 0.7492 & 0.6014 & 0.7322 \\ \hline
\end{tabular}%
\end{table*}

Removing temporal sampling (alongside positive enhancement as it is implemented to balance the negative samples generated by temporal sampling) and negative loop sampling led to performance drops across all monthly datasets, with the only exception being a small improvement in January when negative loop sampling was removed, further confirming that both strategies are meaningful and necessary. For random sender, its removal leads to improvements in January, February, and May datasets, but results in notable declines for other months. Moreover, in January, February, and May, the model loses predictive power for the random sender negative sample type (AUC approximately 0.5). Given the importance of random sender samples in predicting meme stock-related social networks, we conclude that random sender sampling remains both meaningful and necessary.

We conducted an additional ablation study to analyze the effect of positive enhancement. The results are shown in Appendix~\ref{appendix:ab_PE}.

\subsection{Running Time}

\begin{table}[t]
\caption{Running time of one epoch for training different DGE models using different NSSs on the AMC-Jan dataset.}
\label{tab:running-time}
\begin{tabular}{cccc}
\hline
Strategy   & TGNs   & DyGFormer  & GraphMixer  \\ \hline
Random     & 2264 s & 523 s      & 112 s       \\
Historical & 2297 s & 502 s      & 113 s       \\
DINS (proposed)   & 2484 s & 1015 s     & 271 s       \\ \hline
\end{tabular}%
\end{table}

In this subsection, we discuss the additional training time caused by DINS. Although the sampling process itself is not overly complex, the proposed strategy increases the number of negative samples. Therefore, our primary focus is on the training time rather than the complexity of the sampling process. The time needed for one epoch of training three different DGE models using three different NSSs on the AMC-Jan dataset is shown in Table~\ref{tab:running-time}. This metric is selected as: (1) all methods require the same number of batches per epoch, and (2) although the number of involved nodes may vary between batches, the computation time per batch remains consistent when scaling to larger datasets.

For the random and historical sampling strategies, the training times for all models are relatively similar. However, when using DINS, there is an increase. For DyGFormer and GraphMixer, the time required roughly doubles, while for TGNs, the increase is less pronounced, adding only a small amount of additional time. Although the proposed strategy results in a longer training time, especially for DyGFormer and GraphMixer, this increase is compensated by the significant improvement in model performance.

\subsection{Discussion}

Through experiments, we demonstrated the critical impact of NSSs in training and evaluating DGE models for predicting meme stock-related social networks. We validate that our proposed NSSs offers a more comprehensive and accurate evaluation compared to existing methods. Furthermore, we validated that training DGE models using our proposed DINS significantly enhances their ability to predict interactions in meme stock-related social networks. The results showed that, for SOTA DGE model DyGFormer trained using our proposed DINS, the AUC scores consistently reached high levels, with values of at least 0.8 across various datasets. In existing studies, AUC are typically above 0.9 because these models are trained and evaluated with random negative sampling which focus on a single strategy in contrast to our approach which consists of a variety of strategies.

This strong predictive performance indicates that our approach can contribute to a more effective analysis of meme stock-related social networks, which can in turn help in understanding meme stock price movements. Since these networks play a critical role in driving stock price volatility through online discussions and collective sentiment shifts, the ability to accurately model and predict user interactions within these networks provides valuable insights into potential market behavior. By improving the predictive power of DGE models, our proposed strategy could assist in identifying key patterns and interactions that may correlate with significant changes in meme stock prices.

\section{Conclusion}

In this paper, we study the use of DGE models in predicting user interactions on meme stock-related social networks. We analyzed three real-world meme stock-related Reddit networks and demonstrated that the current design of NSSs is insufficient for DGE models to accurately predict interactions. To address this issue, we proposed several individual domain knowledge-informed NSSs and presented a method, DINS, to combine these individual strategies effectively during training. The experiments showed that our proposed NSSs can better evaluate the ability of DGE models in analysing meme stock-related social networks and improve their predictive performance.

Our future work will explore the practical application of DGE models optimized for meme stock-related social network analysis, with the aim of gaining deeper insights into meme stock prices. Besides, since the proposed DINS increases the training time, we will explore integration of active learning techniques to address this downside. Finally, while our study focused on meme stock-related social networks, in our future work, we will explore other types of social networks where DINS could potentially benefit DGE models.

\newpage
\bibliographystyle{ACM-Reference-Format}
\balance
\bibliography{sample-base}

\newpage
\appendix

\section{Summary and Discussion of Proposed NSSs} \label{appendix:summary_NSSs}
Meme stock-related social networks exhibit specific network structures. In this section, we explain how each proposed individual NSS addresses these unique patters: 1) Huge increase in number of interactions over a very short time interval, which requires DGEs to be very effective at predicting when node pairs will interact. Therefore, temporal sampling is designed. 2) Nodes can serve as both source and destination, and frequently switch roles. This makes prediction of interactions between any two nodes necessary. Hence, we designed NSS random receiver and sender. 3) There are loops representing creations of posts which may indicate the users are re-engaging. The NSS negative loop is designed to learn such important properties. 

The proposed DINS can also inherently address rare events and outliers through its targeted sampling strategy. Unlike random sampling that might emphasize outliers, DINS selects negative samples based on temporal and structural patterns in the meme stock related social networks. By leveraging these rather than random selection, our approach naturally reduces the impact of outliers while focusing on meaningful negative samples that better represent the network's evolution. However, quantifying the specific advantages in handling these challenges requires further experimental validation, which we discuss as future work in our conclusion.

\section{Descriptive Statistics of Split Datasets} \label{appendix:statistics_split_datasets}

The descriptive statistics of the monthly split subsets for the BB, AMC, and GME datasets are shown in Table~\ref{tab:sta_sub}. Each row represents the interactions occurring in the training set for a given month, while the validation and test sets contain all interactions that occurred in the following month. Due to the transductive setting, all interactions in the validation and test sets involving nodes that are not present in the training set are removed. The table shows the sizes of the validation and test sets after this adjustment. An exception is made for GME from January to March, where the specific time periods of the interactions included in each subset are detailed in the "Date" column. From Table~\ref{tab:sta_sub}, it can be observed that the proportion of the training set varies across different subsets. This variation allows us to analyze how the size of the training set influences the model's performance on subsequent interactions. 

\section{Hyperparameter Setting} \label{appendix:hypersetting}

In this section, we present the hyperparameter settings of the three DGE methods. The hyperparameters are designed based on the original paper and fine tuned on our datasets. Please refer the original papers \cite{rossi2020temporal,yu2023towards,cong2023we}, for the specific meaning of the hyperparameters.

\paragraph{TGNs}

The batch size is set at 1000, learning rate at 1e-4, memory dimension at 172, number of heads at 2, number of layers at 1, dropout rate at 0.1, number of neighbors at 10, embedding module at graph attention, memory updater at GRU, aggregator at last, message function at identity, and embedding module at graph attention.

\paragraph{DyGFormer}
The batch size is set at 1000, learning rate at 1e-4, channel embedding dimension at 50, patch size at 2, number of layers at 2, number of heads at 2, and dropout rate at 0.1.

\paragraph{GraphMixer}
The batch size is set at 1000, learning rate at 1e-4, number of tokens at 20, number of layers at 2, and dropout rate at 0.1.

\begin{table}[ht]
\caption{Descriptive statistics of the monthly split subsets for the BB, AMC, and GME datasets.}
\label{tab:sta_sub}
\resizebox{\columnwidth}{!}{%
\begin{tabular}{cccccc}
\hline
Month & Training & \begin{tabular}[c]{@{}c@{}}Validation\\ \& Test\end{tabular} & Total     & \begin{tabular}[c]{@{}c@{}}Training Set \\ Percent\end{tabular} & Date                    \\ \hline
\multicolumn{6}{c}{BB}                                                                                                                                                                       \\ \hline
Jan   & 127,634  & 22,041                                                       & 149,675   & 85.27\%                                                              &                         \\
Feb   & 29,342   & 6,208                                                        & 35,550    & 82,53\%                                                              &                         \\
Mar   & 8,479    & 7,033                                                        & 15,512    & 54.66\%                                                              &                         \\
Apr   & 9,197    & 5,671                                                        & 14,868    & 61.85\%                                                              &                         \\
May   & 9,857    & 64,755                                                       & 74,612    & 13.21\%                                                              &                         \\
Jun   & 175,433  & 4,872                                                        & 180,305   & 97.29\%                                                              &                         \\
Jul   & 5,321    & 3,726                                                        & 9,047     & 58.81\%                                                              &                         \\
Aug   & 4,838    & 3,918                                                        & 8,756     & 55.25\%                                                              &                         \\
Sep   & 5,280    & 4,104                                                        & 9,384     & 56.26\%                                                              &                         \\
Oct   & 5,250    & 3,640                                                        & 8,890     & 59.05\%                                                              &                         \\
Nov   & 4,654    & 3,540                                                        & 8,194     & 56.79\%                                                              &                         \\ \hline
\multicolumn{6}{c}{AMC}                                                                                                                                                                      \\ \hline
Jan   & 218,408  & 272,444                                                      & 490,852   & 44,50\%                                                              &                         \\
Feb   & 441,803  & 117,123                                                      & 558,926   & 79,04\%                                                              &                         \\
Mar   & 138,658  & 108,147                                                      & 246,805   & 56,18\%                                                              &                         \\
Apr   & 122,918  & 164,284                                                      & 287,202   & 42,80\%                                                              &                         \\
May   & 215,970  & 276,249                                                      & 492,219   & 43,88\%                                                              &                         \\
Jun   & 363,492  & 135,331                                                      & 498,823   & 72,87\%                                                              &                         \\
Jul   & 143,735  & 110,627                                                      & 254,362   & 56,51\%                                                              &                         \\
Aug   & 120,137  & 66,485                                                       & 186,622   & 64,37\%                                                              &                         \\
Sep   & 73,111   & 43,967                                                       & 117,078   & 62,45\%                                                              &                         \\
Oct   & 48,722   & 41,267                                                       & 89,989    & 54,14\%                                                              &                         \\
Nov   & 45,805   & 38,739                                                       & 84,544    & 54,18\%                                                              &                         \\ \hline
\multicolumn{6}{c}{GME}                                                                                                                                                                      \\ \hline
Sep   & 2,024    & 6,041                                                        & 8,065     & 25,10\%                                                              &                         \\
Oct   & 12,113   & 6,721                                                        & 18,834    & 64,31\%                                                              &                         \\
Nov   & 13,251   & 35,790                                                       & 49,041    & 27,02\%                                                              &                         \\
Dec   & 50,057   & 131,143                                                      & 181,200   & 27,63\%                                                              &                         \\
Jan1  & 243,703  & 558,405                                                      & 802,108   & 30,38\%                                                              & 01.01.2021 - 24.01.2021 \\
Jan2  & 895,344  & 555,120                                                      & 1,450,464 & 61,73\%                                                              & 25.01.2021 - 29.01.2021 \\
Feb1  & 662,931  & 402,929                                                      & 1,065,860 & 62,20\%                                                              & 30.01.2021 - 14.02.2021 \\
Feb2  & 464,310  & 490,503                                                      & 954,813   & 48,63\%                                                              & 15.02.2021 - 28.02.2021 \\
Mar1  & 525,218  & 463,119                                                      & 988,337   & 53,14\%                                                              & 01.03.2021 - 15.03.2021 \\
Mar2  & 474,670  & 275,869                                                      & 750,539   & 63,24\%                                                              & 16.03.2021 - 31.03.2021 \\
Apr   & 290,751  & 33,120                                                       & 323,871   & 89,77\%                                                              &                         \\
May   & 52,559   & 42,196                                                       & 94,755    & 55,47\%                                                              &                         \\
Jun   & 65,572   & 36,135                                                       & 101,707   & 64,47\%                                                              &                         \\
Jul   & 23,839   & 16,830                                                       & 40,669    & 58,62\%                                                              &                         \\ \hline
\end{tabular}%
}
\end{table}

\section{Experimental Results for TGNs and GraphMixer} \label{appendix:results_tgn_graphmixer}

\begin{figure*}[ht]
\centering
\includegraphics[width=\textwidth]{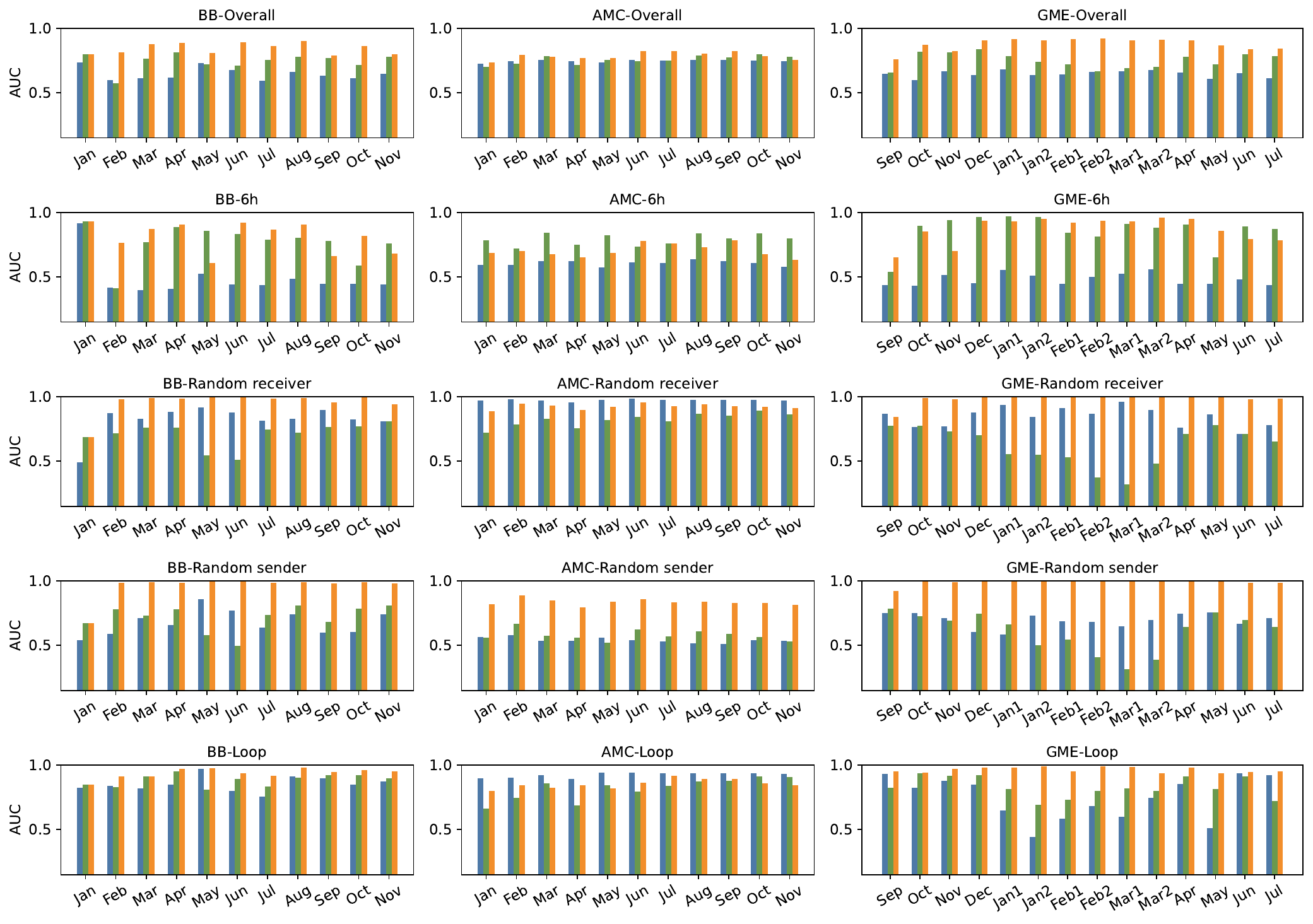}
\caption{Evaluation results of GraphMixer trained with three different NSSs. The \textcolor[HTML]{4E79A7}{blue bar}, \textcolor[HTML]{6A994E}{green bar} and \textcolor[HTML]{F28E2B}{orange bar} (from left to right) show the results of model trained with random NSS, historical NSS and DINS (proposed).}
\label{fig:time-series-GraphMixer}
\end{figure*}

\begin{figure*}[ht]
\centering
\includegraphics[width=\textwidth]{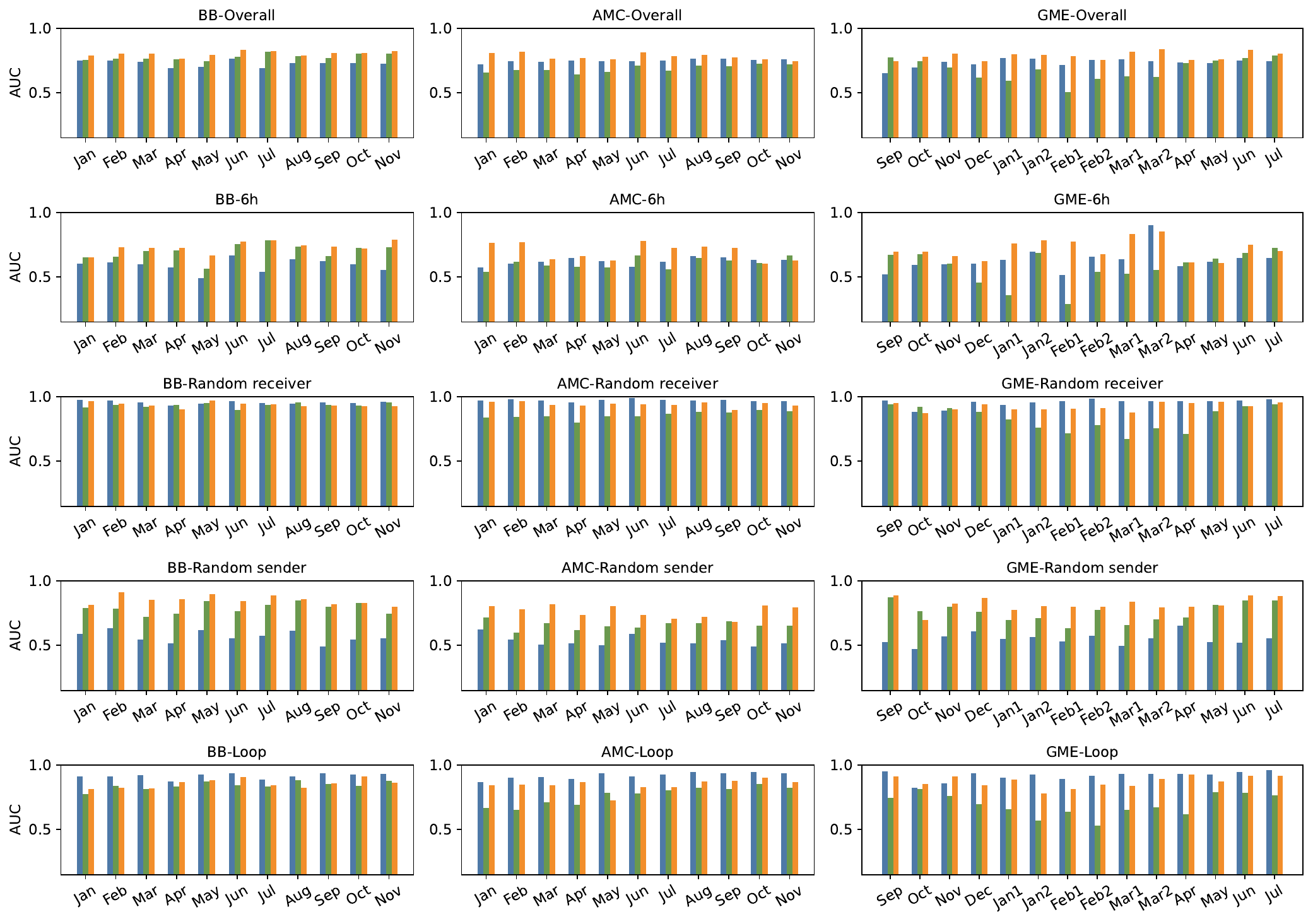}
\caption{Evaluation results of TGNs trained with three different NSSs. The \textcolor[HTML]{4E79A7}{blue bar}, \textcolor[HTML]{6A994E}{green bar} and \textcolor[HTML]{F28E2B}{orange bar} (from left to right) show the results of model trained with random NSS, historical NSS and DINS (proposed).}
\label{fig:time-series-TGNs}
\end{figure*}

In Figure~\ref{fig:time-series-GraphMixer} and Figure~\ref{fig:time-series-TGNs}, we present experimental results of GraphMixer and TGNs on the three datasets, showing how its performance varies with different NSSs used during training and evaluation.

\section{Robustness Experiment} \label{appendix:Robustness}
To ensure the reliability of our results given the large scale of our datasets, particularly GME, we divided the dataset into monthly subsets and conducted experiments on each one, i.e., performing fivefold cross validation.
We conducted additional experiments to provide insight into the variability of the AUC for the proposed method. We trained the three DGEs with our proposed DINS on the BB dataset across all months five times. The average standard deviation of the three DGEs is 0.0087 (DyGFormer), 0.0143 (GrpahMixer) and 0.0076 (TGNs). This demonstrates that the performance of our DINS is consistent across evaluated DGE models. Besides, the randomness is smaller than the margin by which our strategy's AUC exceeds other strategies. 

\section{Ablation Study} 

\subsection{GraphMixer and TGNs} \label{appendix:ab}

In this section, we conduct an ablation study using the BB dataset and GraphMixer and TGNs to validate the impact of individual NSSs. We remove each individual strategy from DINS and train the model with the remaining strategies. The performance is evaluated using overall-type negative samples. The results are shown in Table~\ref{tab:ablation-tgn-graphmixer}.

\subsection{Effect of Positive Enhancement} \label{appendix:ab_PE}

In this subsection, we validate the impact of positive enhancement. We remove positive enhancement from DINS and train the three DGE models on BB dataset. The performance is evaluated using overall-type negative samples. The results are shown in Table~\ref{tab:ablation_PE}.

The results show that the three DGEs trained with ablated DINS (i.e. without positive enhancement) show similar performance in most months compared to the corresponding DGE trained with the complete DINS. In some months, the AUC of DGEs trained with ablated DINS showed a slight decrease. Given that removing temporal sampling and positive enhancement together leads to consistent performance drop (cf. Table~\ref{tab:ablation}), this shows that temporal sampling mainly contributes to the performance improvement, while positive enhancement maintains the performance levels and prevents more substantial performance drops by combating sample imbalance.

\begin{table*}[ht]
\caption{AUC of ablation studies for NSSs for GraphMixer and TGNs on BB dataset.}
\label{tab:ablation-tgn-graphmixer}
\begin{tabular}{l|ccccccccccc}
\hline
Ablation-TGN        & Jan    & Feb    & Mar    & Apr    & May    & Jun    & July   & Aug    & Sep    & Oct    & Nov    \\ \hline
Proposed            & 0.7892 & 0.8060 & 0.8022 & 0.7632 & 0.7949 & 0.8312 & 0.8216 & 0.7914 & 0.8091 & 0.8082 & 0.8218 \\
ablation - future   & 0.7307 & 0.7374 & 0.6966 & 0.7051 & 0.7648 & 0.7660 & 0.7755 & 0.7164 & 0.7746 & 0.7245 & 0.7403 \\
ablation - loop     & 0.7592 & 0.7814 & 0.7850 & 0.7468 & 0.7903 & 0.8332 & 0.8124 & 0.7836 & 0.7984 & 0.8180 & 0.8165 \\
ablation - sender   & 0.7944 & 0.8336 & 0.8197 & 0.8074 & 0.8088 & 0.8116 & 0.8292 & 0.8219 & 0.8228 & 0.8236 & 0.8417 \\ \hline
Ablation-GraphMixer & Jan    & Feb    & Mar    & Apr    & May    & Jun    & July   & Aug    & Sep    & Oct    & Nov    \\ \hline
Proposed            & 0.7096 & 0.7485 & 0.7579 & 0.7683 & 0.7574 & 0.7554 & 0.7721 & 0.7855 & 0.7803 & 0.7435 & 0.7864 \\
ablation - future   & 0.6930 & 0.6576 & 0.7211 & 0.6694 & 0.7112 & 0.7023 & 0.6136 & 0.7000 & 0.5427 & 0.6747 & 0.6734 \\
ablation - loop     & 0.6929 & 0.7503 & 0.7349 & 0.7619 & 0.7197 & 0.7424 & 0.7558 & 0.7788 & 0.7675 & 0.7604 & 0.7794 \\
ablation - sender   & 0.7602 & 0.8271 & 0.8030 & 0.7780 & 0.7752 & 0.7882 & 0.7792 & 0.6875 & 0.7873 & 0.7722 & 0.6149 \\ \hline
\end{tabular}%
\end{table*}

\begin{table*}[ht]
\centering
\caption{AUC change of DGEs trained with DINS without positive enhancement compared with DGEs trained with positive enhancement on BB dataset.}
\begin{tabular}{ccccccccccc}
\hline
Month      & Jan    & Feb    & Mar    & Apr    & May    & Jun    & July   & Aug    & Oct    & Nov    \\ \hline
DyGFormer  & +0.021 & +0.017 & -0.012 & +0.009 & +0.018 & -0.009 & +0.004 & -0.031 & +0.006 & -0.045 \\
GraphMixer & +0.033 & +0.023 & -0.032 & +0.021 & -0.002 & +0.007 & +0.011 & -0.003 & +0.011 & -0.012 \\
TGNs       & +0.019 & +0.015 & -0.027 & -0.004 & +0.005 & +0.013 & +0.012 & -0.024 & -0.001 & -0.019 \\ \hline
\end{tabular}%
\label{tab:ablation_PE}
\end{table*}

\end{document}